\documentclass[aps,prb,superscriptaddress,twocolumn,floatfix,showpacs,a4paper]{revtex4-1}
\usepackage{amsfonts,amssymb}
\usepackage{textcomp}
\usepackage{bm}
\usepackage{upgreek}
\usepackage{graphicx}
\usepackage[latin1]{}
\usepackage{color}
\newcommand{\eps}{\epsilon}
\newcommand{\beq}[1]{\begin{equation} \eqlab{#1}}
	\newcommand{\eeq}{\end{equation}}
\newcommand{\bsub}{\begin{subequations}}
	\newcommand{\esub}{\end{subequations}}
\def\bal#1\eal{\begin{align}#1\end{align}}
\def\bsubal#1\esubal{\bsub \begin{align}#1\end{align} \esub}

\newcommand{\eqlab}[1]{\label{eq:#1}}

\newcommand{\figref}[1]{Figure~\ref{fig:#1}}
\newcommand{\figsref}[2]{Figures~\ref{fig:#1} and~\ref{fig:#2}}

\begin{document} 
	\title{Quantum Transport in Graphene in Presence of Strain-Induced Pseudo-Landau Levels}
 
	\author{Mikkel Settnes}
		\affiliation{ Center for Nanostructured Graphene (CNG), DTU Nanotech  Technical University of Denmark, DK-2800 Kongens Lyngby, Denmark}
		 \affiliation{ Department of Photonics Engineering, Technical University of Denmark, DK-2800 Kongens Lyngby, Denmark} 
	 \author{Nicolas Leconte}
	 	\affiliation{ Catalan Institute of Nanoscience and Nanotechnology (ICN2), CSIC and The Barcelona Institute of Science and Technology, Campus UAB, Bellaterra, 08193 Barcelona, Spain} 
	 	\affiliation{ Department of Physics, University of Seoul, Seoul 130-742, Korea}
	 	\affiliation{ Department of Physics, The University of Texas at Austin, Austin,Texas 78712-1192, USA}	
	 \author{Jose E. Barrios-Vargas} 
	 	\affiliation{ Catalan Institute of Nanoscience and Nanotechnology (ICN2), CSIC and The Barcelona Institute of Science and Technology, Campus UAB, Bellaterra, 08193 Barcelona, Spain}
	 \author{Antti-Pekka Jauho} 
	 	\affiliation{ Center for Nanostructured Graphene (CNG), DTU Nanotech  Technical University of Denmark, DK-2800 Kongens Lyngby, Denmark}
	 \author{Stephan Roche}
	 	\affiliation{ Catalan Institute of Nanoscience and Nanotechnology (ICN2), CSIC and The Barcelona Institute of Science and Technology, Campus UAB, Bellaterra, 08193 Barcelona, Spain}
	\affiliation{ ICREA - Institucio Catalana de Recerca i Estudis Avancats, 08010 Barcelona, Spain}

	
\begin{abstract}
We report on mesoscopic transport fingerprints in disordered graphene caused by strain-field induced pseudomagnetic Landau levels (pLLs). 
Efficient numerical real space calculations of the Kubo formula are performed for an ordered network of nanobubbles in graphene, creating pseudomagnetic fields up to several hundreds of Tesla, values inaccessible by real magnetic fields. Strain-induced pLLs yield enhanced scattering effects across the energy spectrum resulting in lower mean free path and enhanced localization effects. In the vicinity of the zeroth order pLL, we demonstrate an anomalous transport regime, 
where the mean free paths increases with  disorder. We attribute this puzzling behavior to the low-energy sub-lattice polarization induced by the zeroth order pLL, which is unique to pseudomagnetic fields preserving time-reversal symmetry.  
These results, combined with the experimental feasibility of reversible deformation fields, open the way to tailor a metal-insulator transition driven by pseudomagnetic fields.
 \end{abstract}
 \maketitle

Inhomogeneous lattice deformations in graphene generate an effective gauge field modulating the electronic spectrum \cite{Pereira2009,Vozmediano2010,PhysRevB.81.035408}. However, compared to a real magnetic field, the formation of a pseudomagnetic field preserves time reversal symmetry, having an opposite sign in the two inequivalent K and K' valleys \cite{Suzuura2002,PhysRevB.81.161402,Settnes2016}. This leads to different behavior than for real magnetic field especially when introducing disorder. Experimentally, scanning-tunneling measurements on graphene nanobubbles have revealed an electronic spectrum consisting of pseudo-Landau levels (pLL), including a zero-energy peak, showing that moderate spatial deformations can introduce pseudomagnetic field values reaching hundreds of Tesla \cite{Levy2010,Lu2012,Bai2015}. Pseudomagnetic fields (PMF) have also been analyzed in deformed crystals by an atomically controlled arrangement of CO molecules on a gold surface \cite{Gomes2012}, graphene on Ir with intercalated Pb monolayer islands \cite{Calleja2015}, or have been harnessed for designing innovative electronic and photonic graphene devices \cite{Low2010,Schomerus2013,Rechtsman2013,Juan2011,Gradinar2013,Zhu2015}.

However, the \emph{intrinsic} quantum transport fingerprints of graphene in presence of pseudomagnetic fields still needs investigation, especially in disordered systems. In particular, random strain fluctuations are believed to be the dominating disorder source in high-quality on-substrate graphene devices~\cite{PhysRevX.4.041019,Burgos2015}. A signature of the pseudomagnetic $n=0$ pLL state has been predicted for stretched graphene ribbons in the form of a quadruplet low-energy conductance resonance split by edge-induced valley mixing~\cite{Gradinar2013}, but its experimental confirmation remains challenging, and the variable range of transport features in deformed graphene still require further in-depth exploration.

Here we study the effect of pLLs, generated by an ordered network of graphene nanobubbles (or pseudomagnetic dots), using an efficient real space Kubo quantum transport methodology. We consider samples containing electron-hole puddles caused by substrate interactions where the presence of pLLs leads to several anomalous transport features resulting from an intertwined contribution of pseudomagnetic field and disorder effects. The formation of the zero-energy pLL reduces the mean free path by orders of magnitudes in the limit of low defect density ($c$, modeled by a random distribution of Gaussian impurities), but scales as $\ell_{e}\sim c$ up to $c\sim 1\%$, in contradiction with the usual Fermi golden rule argument \cite{Roche2012ji} predicting $\ell_{e}\sim 1/c$ which we obtain by considering the unstrained structure. Additionally, our simulations show that $\ell_{e}\sim 1/\sqrt{B_{s}}$ which evidences a superimposed scaling with the pseudomagnetic length. Finally, we demonstrate further unconventional behavior of the conductivity close to zero-energy where it becomes largely independent of energy for large enough disorder (above $1\%$). Although in the limit of sample length $L\to\infty$, one expects an insulating behavior at zero temperature. These features manifest the strong influence that strain induced pseudomagnetic fields have on the quantum transport properties and also suggest possibilities for metal-insulator transition driven by strain fields.

\begin{figure}[t!]
	\begin{center} 
	 	\includegraphics[width=0.95\columnwidth]{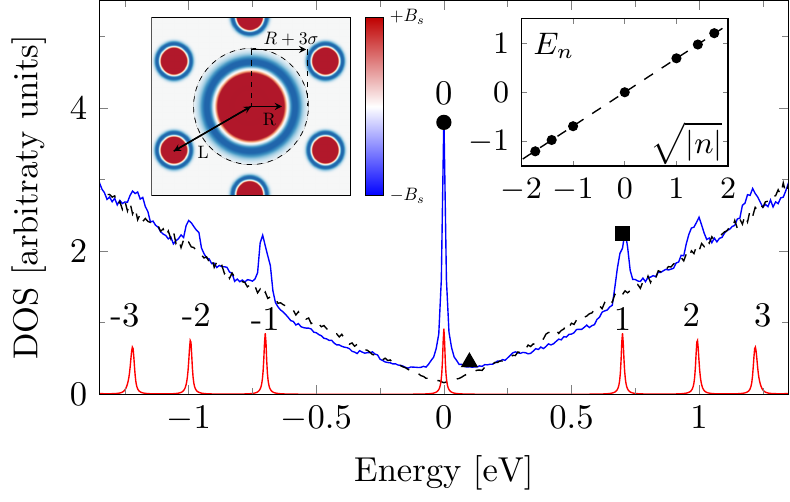}
		\caption{\label{fig:kubo_dos}  \small  DOS for a strain array (blue) with $L=200a $, $R=40a $, $\sigma=10a$ and local strength corresponding to $B_s$ = 450 T (black dashed curve is the unstrained graphene DOS). Both calculations include a $ 0.05\%$ concentration of impurities. The red curve indicate the LDOS in the center of the pseudomagnetic dot averaged over both sublattices for $r<1$ nm (note that the zeroth level resides completely on the B-sublattice \cite{Settnes2015,Neek-Amal2013}). The full DOS (blue) becomes the sum of the unstrained region (dashed, black), the inner part of the strained region (red) and the outer part (not shown).
		The symbols refer to \figref{kubo_D}.
		Left inset shows the superlattice of pseudomagnetic dots with a lattice constant $L$. The central dot has been magnified for visibility illustrating the central region ($r<R$) with a constant pseudomagnetic field surrounded by a region with a pseudomagnetic field of opposite sign $(R<r<R+3\sigma)$.
		Right inset shows the energies of the DOS peaks as a function of the peak number, $\sqrt{n}$, confirming their pLL nature as $E_n = \mathrm{sign}(n) \sqrt{2e\hbar v_F^2B_s|n|}$. 
		}
	\end{center}
\end{figure}
\section{Superlattice of pseudomagnetic dots} To describe the electronic properties of graphene, we use the common tight-binding (TB) model of graphene 

\begin{eqnarray}
\hat{H} = \sum_{i} \epsilon_i c_i^\dagger c_i + \sum_{\langle i,j \rangle} \gamma_0 c_i^\dagger c_j,
\end{eqnarray}
where $\epsilon_i$ is the onsite energy and the sum over $\langle i,j \rangle$ runs over nearest neighbor sites with $\gamma_0=-2.7$ eV. To generate pseudomagnetic fields, we consider a dot with radius $R$ subjected to a planar triaxial displacement, which in polar coordinates are given by \cite{Neek-Amal2013,Guinea2010}

\begin{eqnarray}
{\bf u} = (u_r,u_{\theta}) = (u_0r^2\sin(3\theta),u_0r^2\cos(\theta)), \label{def_field}
\end{eqnarray}
where $r$ and $\theta$ are the polar coordinates and $u_0$ determines the strength of the strain field. This displacement field gives rise to a constant PMF~\cite{Guinea2010} given by $B_s$. Here we ignore out-of-plane and curvature \cite{Settnes2015,Carrillo-Bastos2014,Schneider2015} components of the strain field in order to get a simple connection between the strain and the magnitude of the PMF. This is justified when in-plane strain dominates (to induce pLLs) and no sharp bends are present \cite{Eun-Ah2008,Zenan2014,Pereira2010_PRL}.
The PMF is inherently related to the first order expansion of the TB model, but we emphasize that the calculations reported below do not rely on such approximations, instead we use the modified atomic positions to modify the TB parameters.
Changing the atomic positions according to the displacement field $\bf u$ alters the bond lengths $d_{ij}$, and thereby leads to renormalized TB hopping parameters, 

\begin{eqnarray}
\gamma_{ij} = \gamma_0 \mathrm{e}^{\big[-\beta (a_{ij}/a_0 -1)\big]},\label{strain_gamma}
\end{eqnarray}
where $\beta = \partial \mathrm{log} (\gamma) / \partial \mathrm{log}(a)|_{a=a_0} \approx 3.37$ ~\cite{Pereira2009}.
Here $a_{ij}=\big(a_0^2 + \eps_{xx}x_{ij}^2 + \eps_{yy}y_{ij}^2 + 2\eps_{xy}x_{ij}y_{ij}\big)/a_0$ denotes the modified bond length caused by the deformation, where $a_0=0.142$ nm is the equilibrium bond length and the strain tensor is $\eps_{\nu \mu} = \big(\partial_\mu u_\nu + \partial_\nu u_\mu\big)/2$ with $\nu,\mu=x,y$~\cite{LandauBook}. Using these definitions the deformation field in Equation (\ref{def_field}) gives rise to a magnetic field of $B_s =  8 u_0\hbar\beta/2e a_0$. At last, we note that Equation (\ref{strain_gamma}) can easily incorporate out-of-plane components of the strain as it only depends on the change in bond lengths.
The connection between strain tensor and bond length deformation is only approximate \cite{Midtvedt2016} but sufficient for the present analysis of generic quantum transport fingerprints of strain-induced pLLs. Especially, we note that the effect of the PMF on the LDOS is qualitatively unchanged by relaxation using molecular dynamics methods~\cite{Neek-Amal2013,Zenan2014,Jones2014,Qi2013,Bahamon2015,Neek-Amal2012a,Neek-Amal2012b} and that second nearest neighbor terms only contribute with a scalar potential without generating pseudomagnetic effects \cite{Eun-Ah2008,Neek-Amal2013}.

Outside the central dot region, we apply a smoothing to the strain tensor to assure a soft transition to the strain-free pristine regions. This is accomplished by applying a Gaussian transformation ${\bf \eps} \rightarrow \epsilon' = {\bf \eps}\;\mathrm{exp}[{{-(r-R)^2/2\sigma^2}}]$ for $r>R$, where $R$ is the radius of the PMF region as shown in the left inset of \figref{kubo_dos}. The triaxial strain for $r<R$ gives rise to a constant PMF, whereas for $r>R$ an $r$-dependent PMF of opposite sign develops \cite{Settnes2015}. The field of opposite sign within the smoothing region arises because we apply the smoothing to the physical strain, and {\it not} to the derived PMF.
We repeat this pseudomagnetic dot deformation in a periodic array with lattice constant $L$ (see left inset of \figref{kubo_dos}). This system is an idealized representation of the array of self-formed bubble deformations giving rise to the pLL structure envisioned experimentally by J. Lu \textit{et al.} \cite{Lu2012}. We note that similar calculations have been performed for other array symmetries yielding qualitatively the same conclusions.
 
\section{Kubo transport methodology} 
We study the quantum transport using an order-$N$, real space implementation of the Kubo approach for the conductivity $\sigma_{xx}(E,t)$~\cite{Roche2012ji,Roche1999,Ortmann2011}. The scaling properties of $\sigma_{xx}$ are followed through the dynamics of electronic wavepackets using $\sigma_{xx}(E,t)=e^{2}\rho(E) D_x(E,t)/2$,
where $\rho(E)=\mathrm{Tr} \big[\delta(E-\hat{H})\big]$ is the density of states, $D_x(E,t) = \Delta X^2(E,t)/t$ is the diffusion coefficient and the time- and energy-dependent mean square displacement of the wavepacket is
\begin{eqnarray} 
\Delta X^{2}(E,t) = \mathrm{Tr}\left[\delta(E-\hat{H})\left|\hat{X}(t)-\hat{X}(0)\right|^2\right]/\rho(E),
\end{eqnarray}
where $\hat{X}(t)$ is the position operator in Heisenberg representation. The semi-classical conductivity can be calculated using $\sigma_{xx}^{SC}(E) = e^2 \rho(E) \mathrm{max}_t (D_x(E,t) ) /2$.
Calculations are performed on systems containing approximately $5 \times 10^6$ atoms. 
The state propagation is followed through a total simulation time of $14$ ps and split into three intervals with different time steps $\Delta t_1 = 1$ fs, $\Delta t_2 = 15$ fs and $\Delta t_3 = 60$ fs. The expansion of the time evolution operator uses Chebyshev polynomials corresponding to coefficients larger than $10^{-12}$. Finally, we approximate the traces using random phase states and the Lanczos method with $1500$ iterations and a broadening of $\eta = 5$ meV \cite{RocheBook}.

To mimick the effects of electron-hole puddles induced by the substrate \cite{Adam2009}, the Hamiltonian incorporates long range impurities defined by onsite potential on the $n$'th site $V_n =\sum_i \eps_i \mathrm{e}^{-|{\bf r}_n-{\bf r}_i|^2/(2 \xi^2)}$, where ${\bf r}_i$ is the center of the $i$-th impurity, $\eps_i \in  [-W/2,W/2]$ ($W=2|\gamma_0|$) is the maximum onsite energy and the range is $\xi = 3 a_0$ . 
A low concentration of impurities ($c = 0.05\%$, $c = 0.1\%$ and  $c = 0.2\%$) is distributed randomly throughout the sample allowing us to reach the diffusive regime \cite{Pedersen2014,Ortmann2011,Roche2012ji,Adam2009}.

\begin{figure}
	\begin{center} 
		 \includegraphics[width=0.95\columnwidth]{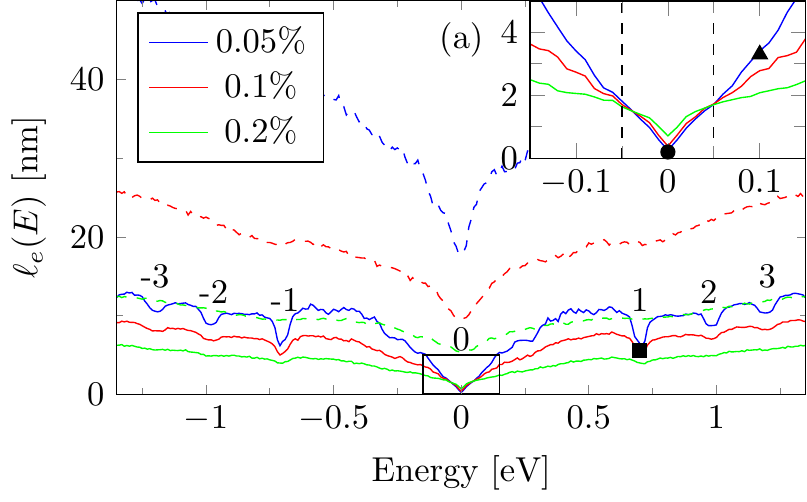}
		 \includegraphics[width=0.95\columnwidth]{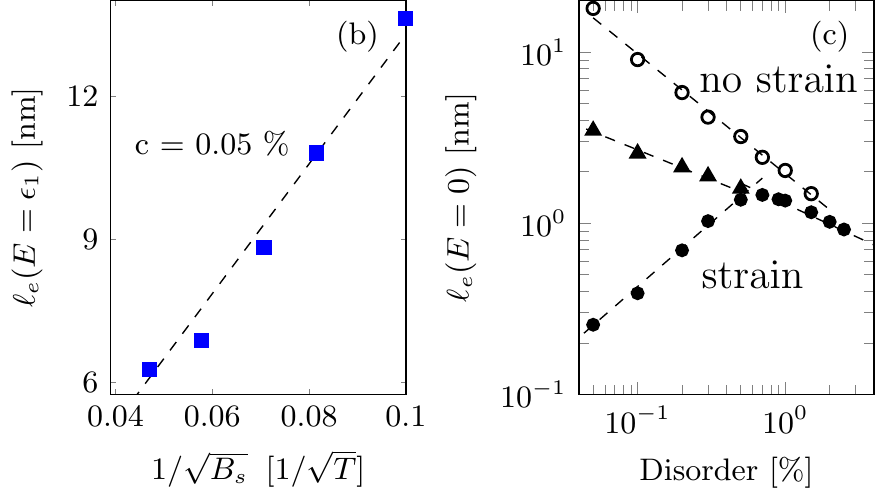} 
		\caption{\label{fig:kubo_mfp}  \small  (a) Mean free path for impurity concentrations $0.05\%$ (blue), $0.1\%$ (red) and $0.2\%$ (green) with (full lines) and without strain (dashed curves). Inset: Zoom of the region around the Diract point as indicated by the black frame. Symbols (circle, triangle and square) refer to \figref{kubo_D}. 
		(b) Mean free path at $c=0.05\%$ for the first pLL at $E=\eps_1$ as a function of the magnitude of the PMF (size of strain).
		(c) $\ell_{e}(E=0)$ as a function of disorder with (filled circle) and without (open circle) strain. For comparison the calculation including strain is also shown for $E=0.1$ eV (filled triangle). Dashed lines are linear fits to different parts of the data.
		}
	\end{center}
\end{figure}

\section{Local density of states and mean free path} 
We first consider the density of states (DOS) of a pseudomagnetic dot array with $L=200a $, $R=40a $, $\sigma=10a$  ($a=\sqrt{3}a_0$), and a maximum strain of approximately $15\%$. The parameters lead to a PMF of $450$ T (see \figref{kubo_dos}), corresponding to a pseudomagnetic length ($\ell_B = \sqrt{\hbar/eB_s} \sim 1.22$ nm). The pseudomagnetic length is smaller than the dot size which is necessary for the appearance of Landau quantization \cite{Settnes2016}. Similar conclusions are obtained for other geometries by varying $L$ and $R$ (not shown). The DOS of the PMF array in \figref{kubo_dos} shows the formation of pLLs following the expected $\sim \sqrt{|n|}$ behavior (\figref{kubo_dos}, right inset), where $n$ is the pLL index. In particular, a strong zero-energy peak is formed, as for real magnetic fields. The peak features are superimposed with the linear dispersion characteristic of the unstrained graphene calculated (black, dashed).

To characterize the impact of pLL states on the quantum transport, we first consider the mean free path (see \figref{kubo_mfp}a), extracted in the diffusive regime as $\ell_{e}(E) =  \mathrm{max}_t (D(E,t) )/{2 v_F}$ where $D(E,t) = D_x(E,t) + D_y(E,t)$. We use $v_F = 8.6\times 10^5$ m/s~\cite{Roche2012ji} even though it should be noted that the Fermi velocity $v_F$ is weakly strain dependent \cite{Juan2011,Manes2013,Ramezani2013,Jang2014,Oliva-Leyva2013,Pereira2010,Pellegrino2010}. In the absence of a strain field (dashed lines), the mean free path follows the Fermi golden rule (FGR) and scales with the impurity density as $\ell_e(E)\sim 1/c$, with values and energy dependence dictated by intervalley scattering \cite{Ortmann2011}.
When superimposing the strain field, we observe a large difference between $\ell_{e}(E)$ with (full lines) and without (dashed lines) the deformation field (see \figref{kubo_mfp}a). For a given impurity concentration $c$, we find a systematic decrease of $\ell_{e}(E)$ compared to the unstrained case. The enhanced scattering induced by strain is particularly significant at the pLL energies where the DOS is large, as revealed by the dips in $\ell_{e}(E)$ in \figref{kubo_mfp}a. At the centers of pLLs, the mean free path scaling is driven by the pseudomagnetic length $\ell_B$ {\it i.e.} $\ell_{e}(E)\sim 1/\sqrt{B_s}$ (see \figref{kubo_mfp}b). 

Next, we focus on the low energy regime in the inset of \figref{kubo_mfp}a. Here, we observe a surprising crossover in the mean free path near $E=\pm 50$ meV (indicated by vertical dashed lines in the inset of \figref{kubo_mfp}a) where the scaling of $\ell_{e}(E)$ with impurity density is reversed going from higher to lower energies. While the high-energy ($|E|>50$ meV) behavior follows the FGR, $\ell_{e}(E)\sim 1/c$ (filled triangles in \figref{kubo_mfp}c), the opposite behavior is observed for $|E|< 50$ meV (filled circles in \figref{kubo_mfp}c). Consequently, we observe an anomalous regime (filled circles) at low energies for which $\ell_e$ increases with disorder. This suggests that the pseudomagnetic field counteracts disorder effects. Increasing the disorder beyond a certain concentration disrupts this mechanism and we recover the more conventional decay of the mean free path with increasing disorder.

\begin{figure}[ t!]
	\begin{center}
		 \includegraphics[width=0.90\columnwidth]{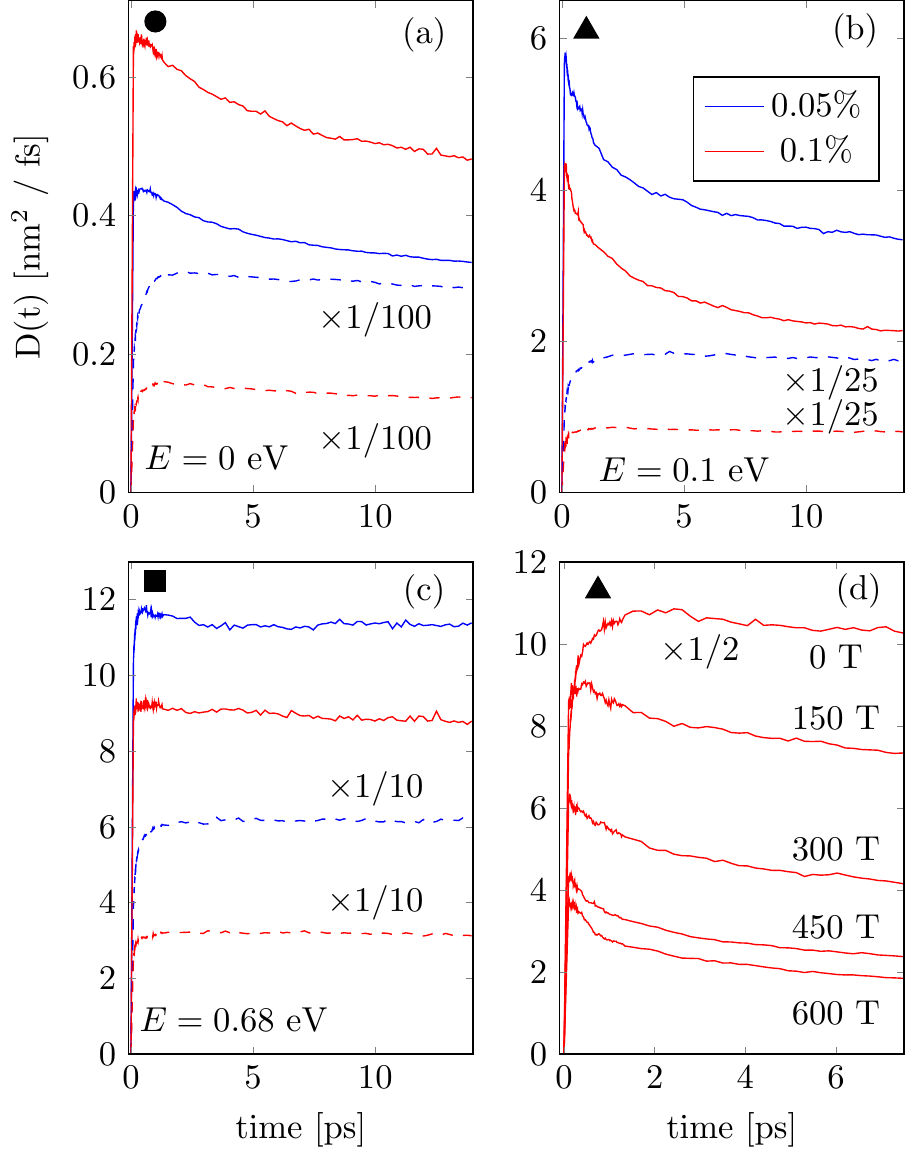}
		\caption{\label{fig:kubo_D}  \small Time-dependent diffusion coefficient with (full curve) and without (dashed curve) strain, for impurity density 0.05\% (blue) and 0.1\% (red) at (a) $E=0.0$ eV, (b) $E=0.1$ eV and (c) $E=0.68$ eV. (d) Same quantity at $E=0.1$ eV for different magnitudes of the strain field corresponding to the indicated PMFs. The symbols (square, circle and triangle) correspond to symbols in \figsref{kubo_dos}{MFP_Plot_zoom}.
		}
	\end{center}
\end{figure}

\section{Localization effects}
To further characterize the transport fingerprints caused by the strain-induced pLLs, we analyze the diffusion coefficients $D(E,t)$ in \figref{kubo_D} at energies marked with symbols in \figsref{kubo_dos}{kubo_mfp}, allowing us to distinguish various transport regimes. A diffusive regime is observed at the energy of the first pLL (\figref{kubo_D}c, square), with a constant asymptotic behavior of $D(E,t)$ both with (full lines) and without a strain field (dashed lines). The additional scattering due to the strain field results in a lower diffusion coefficient leading to the shorter mean free path at the pLL energies as discussed above.

The low energy regime, however, is qualitatively different (\figref{kubo_D}a--b). 
The DOS is dominated by the zeroth pLL, which induces a strong sublattice polarization~\cite{Settnes2016,Neek-Amal2013,Venderbos2015}.
Here the decrease of $D(E,t)$ with time reveals the onset of localization effects. 
In \figref{kubo_D}a--b, we also show the time evolution of the diffusion coefficient in the absence of the strain field (dashed lines). Comparing the curves with and without strain, we conclude that localization effects are generated by the presence of the pseudomagnetic field. This is especially clear in \figref{kubo_D}d showing the variation of $D(E=0.1\;{\rm eV},t)$ while decreasing the pseudomagnetic length. The enhancement of the localization effects are evident through the reduced value and sharper decay of $D(t)$ for higher pseudomagnetic field strengths.

The anomalous transport regime around zero-energy is also manifested in the diffusion coefficients $D(E,t)$. Indeed, $D(E,t)$ exhibits a qualitatively different behavior for $|E|<50$ meV (\figref{kubo_D}a) compared to $|E|>50$ meV (\figref{kubo_D}b) even though the strain-induced pLL causes localization in both regimes. In agreement with the discussion of the mean free path, the absolute value of the diffusion coefficient at low energy increases with impurity concentration. 

\begin{figure}[t!]
	\begin{center}
	 	\includegraphics[width=0.95\columnwidth]{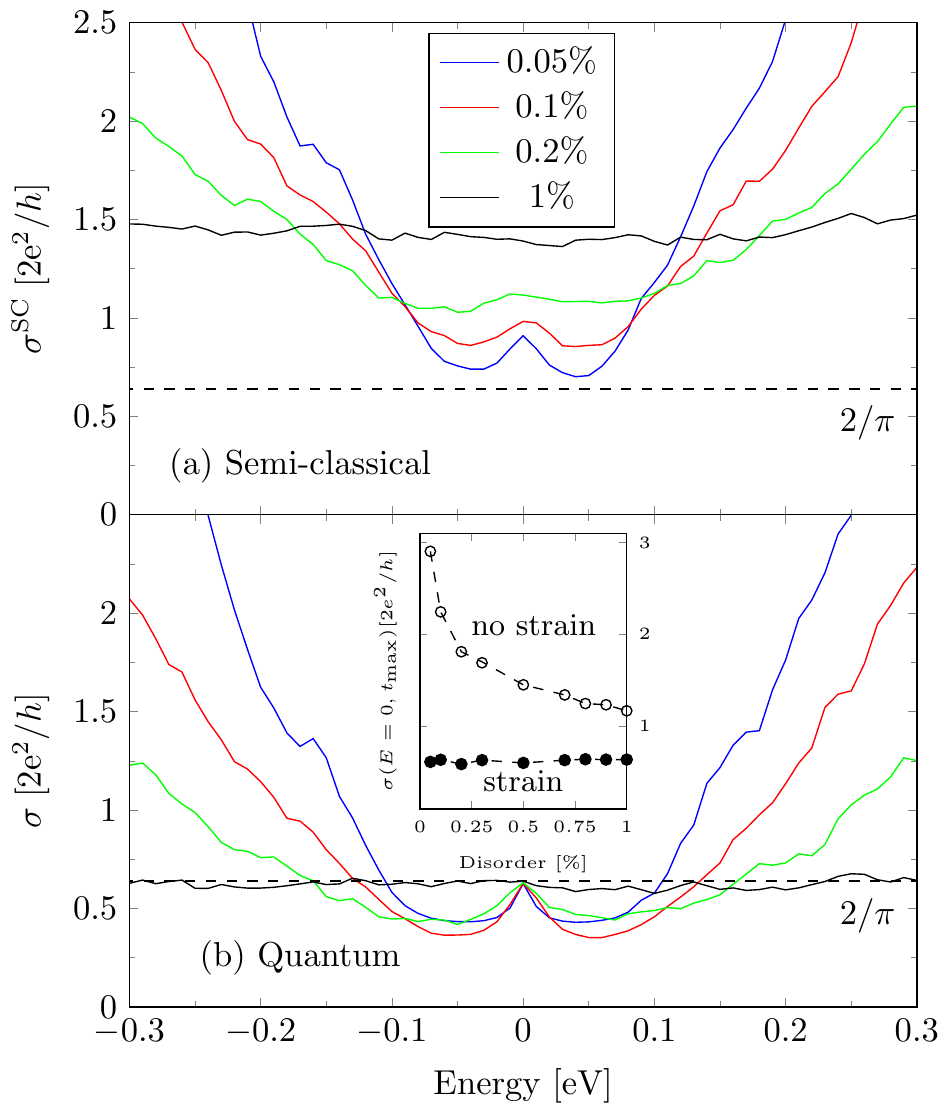}
		\caption{\label{fig:MFP_Plot_zoom} \small (a) Semi-classical $\sigma^{\mathrm{SC}}(E)$ and (b) quantum conductivities, $\sigma(E,t_{max})$, at maximum calculation time for $0.05\%$ (blue), $0.1\%$ (red), $0.2\%$ (green) and $1\%$ (black).  $\sigma_0=\frac{4e^2}{\pi h}$ is indicated by the horizontal solid line. Inset: Quantum conductivity at $E=0$ with (filled circle) and without (open circle) strain as a function of disorder concentration.
		}\label{fig:kubo_sigma}
	\end{center}
\end{figure}

Finally, we consider the semi-classical ($\sigma^{SC}(E)$) and quantum ($\sigma(E)$) conductivities at low energies for different impurity concentrations (\figref{kubo_sigma}a). As seen in the figure, $\sigma^{SC}(E)$ always remains larger than or equal to $\sigma_{0}=4e^{2}/\pi h$ (horizontal dashed line) over the whole energy spectrum, in agreement with the lower bound of the semi-classical value~\cite{Ostrovsky2006,Ostrovsky2010}. The larger value of $\sigma^{SC}(E\sim 0)$ may be related to a low energy behavior, similar to the effect caused by zero-energy modes around lattice monovacancies~\cite{Ferreira2015,Cresti2013,Fan2014}.

The value of $\sigma_{0}$ actually separates two different transport regimes, which are also identified by scrutinizing the long time behavior of $D(E,t)$~\cite{Cresti2013,Fan2014,Lherbier2011,Leconte2011,Trambly2013,Leconte2014}. 
As long as $\sigma(E) > \sigma_{0}$, the system remains diffusive in a metallic ({\it i.e.} non insulating) regime. Correspondingly, the diffusion coefficient saturates at long times. On the other hand, the condition $\sigma(E) < \sigma_{0}$ implies the existence of localization effects, whose strength (weak or strong localization) is dictated by the impurity density and the length-scale (or time-scale) at which the quantum conductivity is evaluated.

The anomalous scaling of the mean free path with disorder is further translated to similar intriguing scaling for the quantum conductivity. In presence of strain, the zero-energy quantum conductivity (taken at the maximum calculation time, $t_{max}$) remains very close to the fundamental semi-classical limit $\sigma_0$ (see \figref{kubo_sigma}b-inset). Similarly to the semi-classical conductivity (\figref{kubo_sigma}a), the dynamical conductivity, $\sigma(E,t_{max})$, shows a complex energy-dependent profile evolution with defect concentration.  
According to \figref{kubo_sigma}b, the conductivity for $|E| \geq 0.1$ eV decays with defect density, {\it i.e.} $\sigma(|E|\geq 0.1{\rm \; eV},t_{max})\sim 1/c$. However, at lower energies such behavior is interrupted and $\sigma(|E|\leq 0.1{\rm \; eV},t_{max})$ instead increases with defect density. A special situation is observed in \figref{kubo_sigma}b-inset where $\sigma(|E|\sim 0 {\rm \; eV},t_{max})$ is almost constant for a wide range of disorder. 

The asymmetric disorder effect between low and high energy is in sharp contrast to many types of disordered graphene systems with similar mean free paths (namely on the order of a few nanometers) at zero magnetic field \cite{Cresti2013,Fan2014,Lherbier2011,Leconte2011,Trambly2013,Leconte2014}, and resembles the percolation mechanisms ~\cite{ Leconte2014,Thouless1981,Halperin1982,Prange1982b,Iordansky1982,Floser2013,Kazarinov1982,Luryi1983,Trugman1983,Giuliani1983,Tsukada1976, PhysRevB.93.115404}, suggesting that the zeroth order pLL plays a critical role. This requires additional insight, provided in the next Section.
\section{Energy-dependent random disorder effect}
\label{disorderDependency}

In this section, we provide a tentative interpretation for the opposite effect of the Gaussian disorder on the low and high-energy behavior, in the classical transport regime (dictated by the mean free path). To this end, we scrutinize the role of real-space distribution of electronic states subjected to a pseudomagnetic field. It has been shown that low-energy states are preferentially located inside the deformed region, while high-energy states correspond to states outside the bubbles~\cite{Moldovan2013}. Furthermore, the zeroth pLL induces sublattice polarization \cite{Neek-Amal2013}, where the preferred sublattice depends on the deformation direction \cite{Settnes2016}. This sublattice polarization of the first Landau level is a unique feature of pseudomagnetic fields. The strain direction considered here causes the $n=0$ pLL states in the central region (red color in left inset of Fig.~\ref{fig:kubo_dos}) to localize on the B-sublattice. On the other hand, the states in the outer ring (blue color) with an opposite pseudomagnetic field sign, is A-sublattice polarized. Consequently, the low-energy states inside the bubbles tend to be confined by this opposite sublattice polarization from the normal conducting states outside the bubbles in the undeformed regions. Now, because we introduce long-range disorder randomly on both A and B sublattices everywhere in the sample, the increased disorder concentration disturbs the perfect low-energy sublattice polarization. Thus, the spatial sublattice-confinement of the states is broken, leading to an increased mean free path and conductivity. 
We note that this scenario is compatible with the fact that the quantum localization effects increase with disorder, as expected (see for instance Fig.~\ref{fig:kubo_D}a).

\begin{figure}[t]
	\begin{center}
		
		\includegraphics[width=0.75\columnwidth]{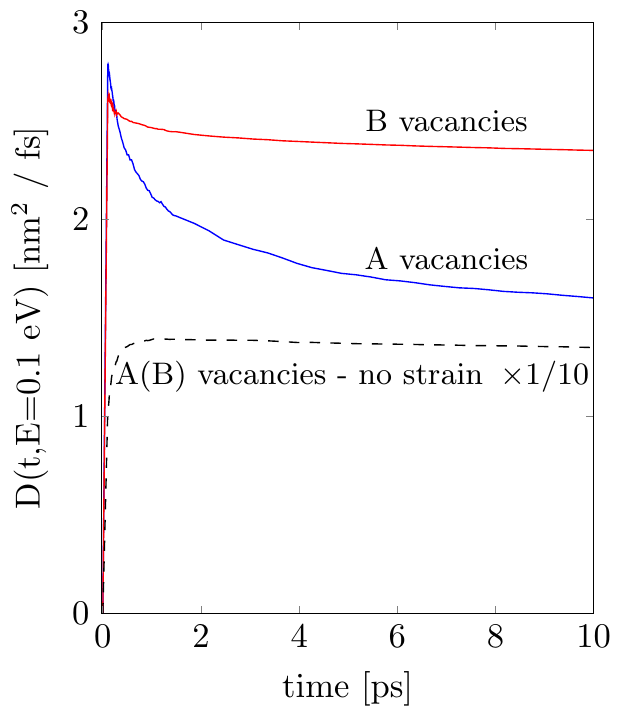}
		
		\caption{\label{fig:kubo_vacancy} \small Diffusion coefficients at $E=0.1$ eV for 0.3 \% vacancies either distributed on the A (blue) or B (red) sublattice. Without strain (dashed curve) the diffusion coefficient for A or B vacancies falls on top of each other. 
		}\label{fig:kubo_sigma_L}
	\end{center}
\end{figure}


To further confirm the sublattice-polarization scenario, we introduce a sublattice-selective (vacancy) disorder, which has been shown to provoke asymmetric transport phenomena~\cite{Lherbier2013,Leconte2011_acs}. We model such unreconstructed vacancies by removing carbon atoms randomly everywhere in the sample. In \figref{kubo_vacancy}, the diffusion coefficients for sublattice-selective vacancy disorder, located either on the A- or B-sublattice, are shown, both with and without strain. The inequivalent effect of A- and B-vacancies exists only when the pseudomagnetic field is present. 
 Such vacancies increase the number of states in the opposite sublattice from where they reside. As such, the A-vacancies increase the number of states in the B-sublattice. Because the low-energy states are mostly polarized on the B-sublattice, they are strongly affected by short-range valley-mixing effects induced by A-vacancies, leading to stronger localization effects (blue curve on \figref{kubo_sigma_L}). Indeed, unlike the zeroth order Landau level caused by a real magnetic field where valleys are located on opposite sublattices, the valleys are concentrated on the same sublattice in the presence of a pseudomagnetic field \cite{Settnes2016}. Finally, even if the A-vacancies yield stronger localization effects, they concurrently lead to longer mean-free paths, when compared to B-vacancies. This observation further supports our interpretation since A-vacancies (increased B-sublattice polarization) increase the number of states to percolate through the blocking A-polarized ring around the bubbles.



\section{Conclusion} 
The presence of strain-induced pseudomagnetic fields forming pseudo-Landau levels has been shown to trigger unconventional quantum transport features compared to other types of disordered graphene systems. The transport physics close to half-filling is particularly remarkable, with a huge drop of the transport mean free path upon switching on a pseudomagnetic field, indicating the possibility of a mechanically induced metal-insulator transition. 

We note, however, that despite of the weak localization caused by the deformation field, the strength of quantum interferences, even in the large disorder limit, remains anomalously weak for the sublattice polarized zeroth pseudo Landau level (compared to other types of disorder). This anomalous scaling with disorder gives a peculiar transport fingerprint of the combination of pseudomagnetic field   and disorder effects, especially for the sublattice-polarization unique to the pseudomagnetic field compared to real magnetic fields.

Superlattice networks of deformation fields have been experimentally demonstrated \cite{Reserbat2014}, and could be tunable under pressure \cite{Zenan2014,Bunch2008,Khestanova2016} or by temperature. In this way, varying the mean free path upon deformation, could reversibly switch the system between metallic and insulating states (for $\xi/L\ll 1$, with $\xi$ the localization length and $L$ the sample size). Finally, the role of pseudomagnetic field in spin transport deserves further consideration, in the context of spin manipulation and graphene spin-based devices  \cite{Roche2015}.

	\textbf{Acknowledgement:}
	The work by M.S. is funded by the Danish Council for Independent Research (DFF – 5051-00011). Funding from the European Union Seventh Framework Programme under grant agreement 604391 Graphene Flagship is acknowledged.  S.R. acknowledges the Spanish Ministry of Economy and Competitiveness for funding (MAT2012-33911), the  Secretaria de Universidades e Investigacion del Departamento de Economia y Conocimiento de la Generalidad de Catalu\~{n}a and the Severo Ochoa Program (MINECO SEV-2013-0295). The Center for Nanostructured Graphene (CNG) is sponsored by the Danish National Research Foundation (DNRF103).

\providecommand{\newblock}{}

\end{document}